\documentclass{mystyle}

\usepackage{graphicx}

\usepackage{amsmath}  
\usepackage{amssymb}  

\newcommand{\g}{$\gamma$} 
\newcommand{\F}{\textit{Fermi}} 

\title{Science highlights from the \F{} Large Area Telescope}

\author{L.~Baldini\from{ins:infnpisa} \atque L.~Tibaldo\from{ins:unipadova}
  on behalf of the \F{} LAT collaboration}
\instlist{%
  \inst{ins:infnpisa} INFN-Sezione di Pisa
  \inst{ins:unipadova} Dipartimento di Fisica ``G. Galilei'',
  Universit\`a di Padova, and INFN-Sezione di Padova
}
\PACSes{\PACSit{95.85.Pw}{Astronomical \g-ray observations}
\PACSit{98.70.Sa}{Cosmic rays}
\PACSit{95.35.+d}{Dark matter}}

\begin{document}

\maketitle

\begin{abstract}
  During its first three years of operation, the \textit{Fermi \g-ray
    Space Telescope} has provided an unprecedented view of the high energy
  \g-ray sky, and also performed direct measurements of the cosmic-ray
  leptons and searches for signals from dark matter. In this paper we present
  a short overview of some highlight results, shedding new light on the
  high-energy side of the Universe. 
\end{abstract}

\section{The \textit{Fermi \g-ray Space Telescope}: an observatory of the high-energy Universe}

Radiation in the \g-ray domain is a privileged messenger of high-energy processes
taking place in
our Universe. Contrary to charged cosmic rays (CRs), which are deflected by magnetic fields and
rapidly
loose memory of their sources, \g-rays carry directional information. Compared to neutrinos or
gravitational waves they are
easier to detect thanks to their larger interaction probabilities.

Designed to survey the \g-ray sky in the broad energy range from 20~MeV to
more than 300~GeV, with the additional capabilities of studying transient
phenomena at lower energies and charged species, notably leptons, from GeV to TeV energies, the
\textit{Fermi \g-ray Space Telescope}\footnote{Formerly \textit{\g-ray Large Area Space Telescope},
 \textit{GLAST}.} is the premier space-borne \g-ray observatory of this
decade.

\F{} carries two instruments on-board: the \g-ray Burst
Monitor (GBM)~\cite{ref:gbmpaper} and the Large Area Telescope
(LAT)~\cite{ref:latpaper}.
The GBM, sensitive in the energy range between 8~keV and 40~MeV, is designed
to observe the full unocculted sky with rough directional
capabilities (at the level of one to a few degrees) for the study of transient
sources, particularly \g-Ray Bursts (GRBs).
The LAT is a pair conversion telescope for photons above 20~MeV up to
a few hundreds of GeV.

\subsection{The LAT instrument and its performance}

The LAT is a $4 \times 4$ array of identical towers, each one made by a
tracker-converter module (hereafter \emph{tracker}) and a calorimeter module. A
segmented anti-coincidence detector (ACD) covers the tracker array
and a programmable trigger and data acquisition system completes the instrument.
Though owing most of the basic design to its predecessors---particularly the
Energetic \g-Ray Experiment Telescope (EGRET)~\cite{ref:egretpaper} on-board the
\textit{CGRO} mission---the LAT exploits the state of the art in terms of detector
technology, which allows for a breakthrough leap in the instrument performance.

Each tracker module features 16~tungsten layers, promoting the conversion of
\g-rays into $e^+/e^-$ pairs, and 18~$x$-$y$ pairs of single-sided silicon
strip detector planes---for a total of 1.5~radiation lengths of material on-axis.
The silicon-sensor technology allows precise tracking (with no detector-induced
dead time and no use of consumables) and the capability to self-trigger.

Each calorimeter module consists of 96~CsI(Tl) crystals, arranged in
a hodoscopic configuration (for a total depth of $\sim 8.6$~radiation lengths
on axis).
The calorimeter provides an intrinsically three-dimensional image of the shower
development, which is crucial both for the energy reconstruction (especially
at high-energy, where a significant part of the shower can leak out of the back
of the instrument) and for background rejection.

The anti-coincidence detector, a set of plastic scintillators surrounding the tracker,
is the first defense of the LAT against the overwhelming background due to charged CRs.
In order to limit the ``self-veto'' effect---due to the back-splash of secondaries
from high-energy particles hitting the calorimeter---it is segmented in 89 tiles 
providing spatial information that can be correlated with the signal from the
tracker and the calorimeter.

The design, construction and operation of such a complex detector is a
fascinating subject on its own and the interested readers can refer to~\cite{ref:latpaper}
and references therein for further details. The LAT largely surpasses the
previous generations of \g-ray telescopes in terms of effective area,
energy range, instrumental dead time, angular resolution and field of view (2.5~sr).
It has the ability to observe 20\% of the sky at any time which, in the
nominal scanning mode of operation, enables it to view the entire sky
every three
hours.

\subsection{Outline}

With the observatory being well into the third year of sky survey,
this paper is a short overview of the most important Science highlights,
chosen in consideration of the interests of the audience and presented by
the authors in two review talks at the XXV Rencontres de
Physique de La Vall\'ee d'Aoste.

Section~\ref{sec:gamma} presents a short overview of the high-energy \g-ray sky unveiled by the
\F{} observatory. The different \g-ray sources are examined to understand the non-thermal processes
leading to the production of high-energy particles, and also to view them in relation to other
messengers such as charged CRs, neutrinos and gravitational waves. Then, Section~\ref{sec:dm} is
devoted to the search
for dark matter (DM) signals in \g-rays with the LAT. Finally, Section~\ref{sec:cr} summarizes the
experimental advances due to the LAT in the measurement of the leptonic component of the cosmic
radiation, and its implications in the light of results from other experiments.

\section{The high-energy \g-ray sky seen by \F}\label{sec:gamma}

The sky in high-energy \g-rays is dominated
by diffuse emission: more than 70\% of the photons
detected by the LAT are produced in the interstellar space of our Galaxy by interactions of
high-energy CRs with matter and low-energy radiation fields. An additional diffuse component with
an almost isotropic distribution, and therefore thought to be extragalactic in origin, accounts for
another sizable fraction of the LAT photon sample. The rest consists of sources detected by the LAT
from a variegate zoo of objects
including Active Galactic Nuclei (AGN) and normal galaxies, pulsars and their relativistic wind
nebulae, globular clusters, binary systems, shock-waves remaining from supernova explosions and
nearby solar-system bodies like the Sun and the Moon.

\begin{figure}[!bp]
\begin{center}
\includegraphics[width=0.7\textwidth]{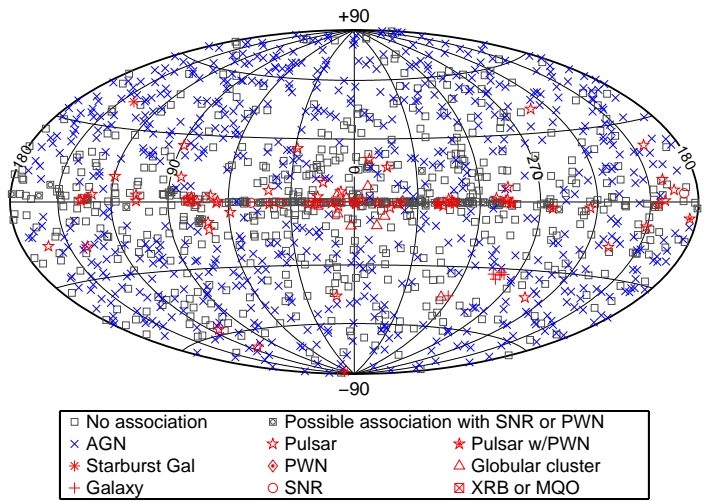}
\caption{Skymap showing the positions of the 1451 1FGL sources \cite{ref:1FGL}
(Aitoff projection in Galactic coordinates). The different markers correspond to
the association at other wavelengths most likely for each source.}\label{fig:1FGL}
\end{center}
\end{figure}

A major step forward in the understanding of the \g-ray sky was given by the first LAT source
Catalog (1FGL) \cite{ref:1FGL}, based on the first 11~months of data, which is summarized by
Fig.~\ref{fig:1FGL}.
The LAT increased the number of known \g-ray sources from a few hundred to 1451, including new
classes of objects like globular clusters and starburst galaxies. The mystery of the unassociated
\g-ray sources, which comprise 630 out of the 1451 sources in the 1FGL sample, continues to puzzle
astrophysicists. The second LAT source Catalog, based on two years of data and improved detection
methods, will be released soon.

Compared to what is known at longer wavelengths, a characteristic feature of the \g-ray sky is its
rapid variability over timescales from a few seconds to months.
In addition to
transients known for decades like \g-ray bursts and flares from AGN, the LAT observed
unexpected phenomena like \g-ray emission from the nova in the symbiotic binary V407 Cygni
\cite{ref:Nova} and variability from the Crab nebula \cite{ref:variableCrab}.

\subsection{\g-ray bursts}
The mysterious explosions known as \g-ray bursts (GRBs), which episodically
outshine for a few seconds to minutes any other sources in the \g-ray sky, have intrigued scientists
since
their discovery in the '60s. There is currently a large consensus that
they are produced by a release of gravitational energy (of the order of one rest solar mass)
over a very short time interval (of the order of 1~s) within a very compact region (of the
order of
10~km) by a  cataclysmic event, either the collapse of a
massive-star core into a black hole (long GRBs) or by the merging of two compact objects (short
GRBs).
This energy release would mostly result in a burst of thermal neutrinos and perhaps gravitational
waves, but also in a very high temperature fireball expanding at highly relativistic speed which
would 
undergo energy dissipation, producing, among other particles, \g-rays and later developing into a
blastwave which would decelerate against the external medium and give rise to the afterglow
observed in
\g-rays and at longer wavelengths. The explosions producing GRBs are among the few phenomena
sufficiently energetic to contribute to the acceleration of CRs across the whole observed
energy spectrum.

Observations indicate that GRBs are distributed isotropically in the sky and that those which
have known redshift are located at cosmological distances. \F, with the LAT and the GBM,
observes GRBs over 6~orders of magnitude in photon energy: it was therefore expected to
shed light on some fundamental aspects of GRB Physics, including the origin of the energy and the
mechanism by which it is transported, the
\g-ray emission mechanism and
the level of collimation (isotropic emission would require an unrealistic energy release). The
GBM has, so far, observed a few hundred GRBs, of which a few tens were detected
also
by the LAT. The improved performance of the LAT, especially at energies $>10$~GeV, led,
for some bursts, to the measurement of a hard
spectral component in addition to the standard Band component, e.g. \cite{ref:GRB090902B}, or of
high-energy afterglows, e.g. \cite{ref:depasquale}. 
 
The highest energy photon ever recorded from a burst was detected at $\sim 31$~GeV  from GRB~090510
\cite{ref:GRB090510nature} at a redshift $z=0.903$. In simple radiation models such as that
described in \cite{ref:GRB090510apj}, the temporal structure of the
burst requires, in order to lower the internal opacity due to \g-\g{}
interactions, bulk Lorenz factors larger than a few hundreds. High-energy photons traveling over
cosmological distances are also a powerful tool to probe for a possible breaking of Lorenz
invariance. Assuming that a photon of energy $E$ is delayed by $\Delta t=E/(M_\mathrm{LIV}\,c^2)$,
where
$M_\mathrm{LIV}$ sets the scale of the Lorenz invariance violation, the temporal structure of
GRB~090510 implies $M_\mathrm{LIV}>1.19\,M_\mathrm{Planck}$ \cite{ref:GRB090510nature}.

\subsection{Active Galactic Nuclei}\label{agn}
The largest class of associated sources in the LAT Catalog corresponds to AGN
\cite{ref:1FGL,ref:1LAC}. High-energy emission from AGN is thought to be powered by the accretion
of matter onto a super-massive black hole. In a process not fully understood yet, this produces a
jet of relativistic particles that shoots away from the central engine. There are, however,
pending questions, including the emission mechanism which produces the observed \g-ray emission and
the region where this process takes place. There is not even a consensus on the nature of the
particles carrying the energy which is radiated in \g-rays, either leptons or nucleons. In the
latter case AGN are promising neutrino sources. There is also a suggestive correlation measured by
the Auger observatory between the arrival directions of ultra-high energy
CRs and nearby AGN \cite{ref:Auger}, which might be their primary sources.

Thanks to the improved angular resolution of the LAT, Centaurus~A has become
the first AGN ever
resolved in high-energy \g-rays; the \g-ray flux is almost equally divided between the core of
the Galaxy and the lobes, which are flooded by electrons of energies up to 1~TeV, accelerated
directly in the lobes or efficiently transported from the core \cite{ref:CenA}.

 A \g-ray flare from the AGN 3C279, associated with a change in the optical polarization angle,
proved that, in this object, the regions of \g-ray and optical emission are co-located; the change
in
optical polarization implies a non-axisymmetric magnetic field and a distance of the emitting
material from the central engine $>10^5$ gravitational radii \cite{ref:3C279}.

\subsection{Pulsars and pulsar wind nebulae}
Pulsars, i.e. highly-magnetized rotating neutron stars emitting periodic spikes of electromagnetic
radiation, were the first class of identified \g-ray sources. The LAT brought the number of known
\g-ray pulsars from 6 to $>70$ and further new discoveries are foreseen. In addition to young radio
pulsars
discovered in \g-rays thanks to the rotational properties measured by radio astronomers, the LAT
discovered two new pulsar populations:
\begin{itemize}
 \item \g-ray selected pulsars found without any prior knowledge about their pulsation
\cite{ref:blindPSR}; some of them did not reveal, so far, any radio emission, suggesting that the
\g-ray beams may be broader than those at longer wavelengths;
 \item millisecond pulsars \cite{ref:msPSR}, i.e. older pulsars with weaker surface magnetic fields,
which were spun up to higher velocities by accreting material from a companion object in a binary
system. 
\end{itemize}
There is a large consensus that the radiation is mostly due to interactions of leptons accelerated
at
large magnetic-field strengths. The properties illustrated in the first LAT pulsar Catalog
\cite{ref:PSRcat} suggest that \g-ray emission largely arises from the outer magnetosphere of the
neutron stars rather than the region near the magnetic poles as previously proposed by several
authors.

Most of the energy lost by pulsars while spinning down goes into the production of a pulsar wind,
and its termination shock further accelerate particles, primarily electrons, which produce a pulsar
wind nebula (PWN). Pulsars and their wind nebulae might therefore significantly contribute to the
leptonic fraction observed in cosmic rays. The enhanced sensitivity of the LAT enabled, so far, the
detection of \g-ray emission from three PWNe, namely the Crab Nebula, Vela~X and MSH~$15-52$
\cite{ref:PWNcat}. The LAT detected a surprising day-scale variability from the Crab nebula
\cite{ref:variableCrab}: this was interpreted as synchrotron emission from an electron population
reaching PeV energies in a region within 17 light days of the neutron star, i.e. at the
termination  shock of the pulsar wind.

\subsection{Supernova remnants}

Shock-waves produced by supernova explosions have long been considered the most likely candidates
for the
parent source population of Galactic CRs, since, except for GRBs, they are the only objects
sufficiently energetic to
explain the CR fluxes observed at the Earth. Diffusive shock acceleration in supernova remnants
(SNR) provides a plausible mechanism to sustain Galactic CRs. While multiwavelength spectra of SNRs
undoubtedly prove the presence of high-energy electrons in the expanding shock-waves, only indirect
evidence points to the acceleration of nuclei. In the future, the detection of high-energy
neutrinos might be the smoking gun of hadron acceleration.

\begin{figure}[!btp]
\begin{center}
\includegraphics[width=1.\textwidth]{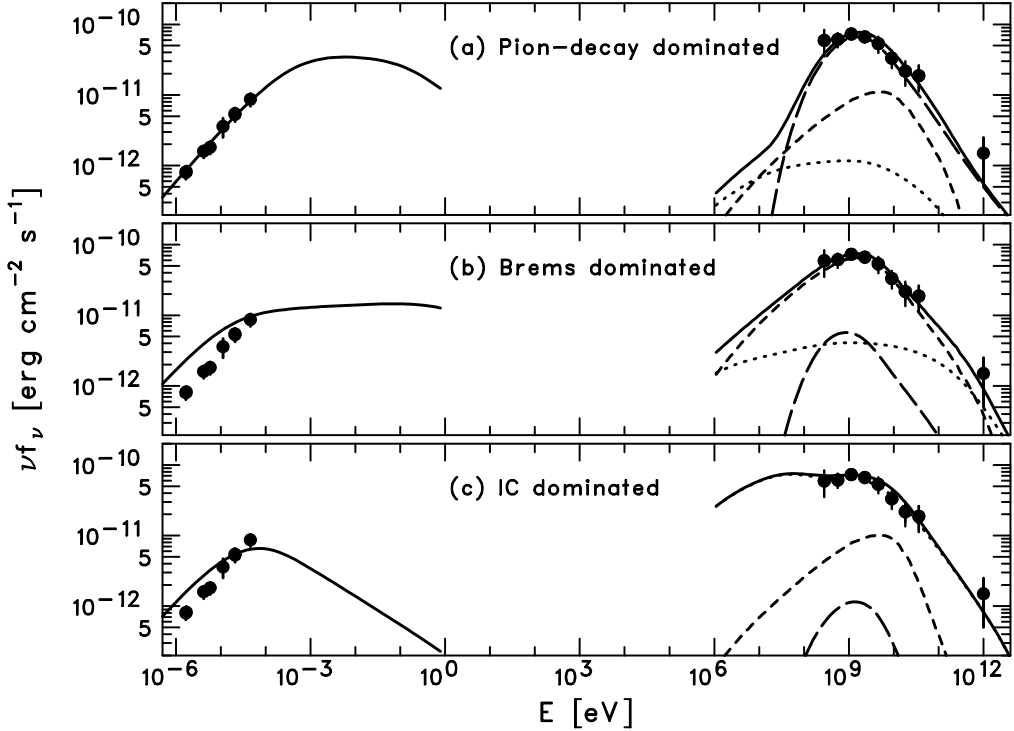}
\caption{Broad-band spectral energy distribution of SNR W51 compared with different emission models
\cite{ref:W51}. The spectrum combines LAT measurements from 200~MeV to 50 GeV, with
TeV measurements by the HESS telescope and radio measurements of synchrotron
emission from relativistic electrons. Model (a) corresponds to the case of efficient
nuclei acceleration (where the \g-ray emission is dominated by $\pi^0$ decay), models (b) and (c)
two different cases of inefficient nuclei acceleration, where \g-ray emission is dominated by
Bremmstrahlung and inverse-Compton scattering, respectively.}\label{fig:W51}
\end{center}
\end{figure}

\g-ray observations are an important probe of non-thermal particles in SNR shock-waves. The improved
angular resolution of the LAT enabled us to spatially resolve
some SNRs, e.g. W44 \cite{ref:W44}, and to
verify that \g-ray emission is associated with the shock region. The dominant class of LAT detected
SNRs, like W44, are middle-aged SNRs (with ages of the order of $10^4$ years) showing interactions
with molecular clouds. Their broad-band spectrum is better reproduced by assuming that the
\g-radiation is mostly produced by nucleon-nucleon inelastic collisions, as shown for W51 in
Fig.~\ref{fig:W51}.
Middle-aged SNRs detected
by LAT show a steep spectrum above a few GeV; on the other hand young SNRs (ages of a few
thousand years), where efficient particle acceleration is thought to occur, show in the LAT energy
band hard spectra which connect with the TeV emission detected by ground-based instruments. 

\subsection{Galactic interstellar emission\label{sec:gde}}
Galactic interstellar emission is produced by CR interactions, via nucleon-nucleon inelastic
collisions (through $\pi^0$ production and decay) and via electron Bremmstrahlung and
inverse-Compton scattering. It is therefore, together with synchrotron radiation from electrons at
radio wavelengths, the only probe of CRs on the Galactic scale, beyond direct measurements performed
in the solar system and at its outer frontiers. Previous measurements showed an excess at
energies $\gtrsim 1$~GeV with respect to expectations based on the directly measured CR spectra
\cite{ref:strong2000}, attributed either to CR spectral variations, instrumental effects or
contributions from exotic phenomena.

The LAT measured the \g-ray emission from the nearby interstellar space and found that its spectrum
is softer than previous measurements \cite{ref:nogevexc}, in good agreement within 10\% with
production by CRs
with a spectrum consistent with that directly measured near the Earth
\cite{ref:nogevexc,ref:hiemiss}. The LAT is currently investigating, through interstellar emission
observations, the distribution of CRs at large in the Galaxy. LAT observations
reinforced the so-called CR gradient problem, with the measurement of CR densities larger than
expected toward the outer Galaxy, where the putative sources, SNRs, rapidly drop off
\cite{ref:iiquad,ref:iiiquad}; this might be related to some poorly-constrained aspects in the
propagation of particles (e.g. a large diffusion halo, anisotropic diffusion or convection) or to
large masses of target gas escaping our observations or, less likely, to unknown accelerators. The
ongoing modeling of large-scale properties of interstellar \g-ray emission in the light of LAT data
\cite{ref:strong2011} and the studies of other selected region will deepen our understanding of
particle propagation in the Galaxy in the near future.

\begin{figure}[!hbtp]
\begin{center}
\includegraphics[width=0.7\textwidth]{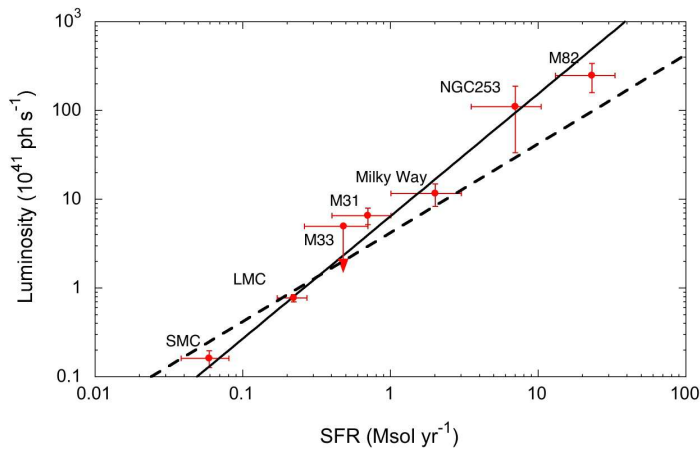}
\caption{\g-ray luminosity as a function of star-formation rate for some
galaxies observed by the LAT \cite{ref:m31}.}\label{fig:galaxies}
\end{center}
\end{figure}

\subsection{External galaxies}\label{extgal}

High-energy \g-ray emission from non-active external galaxies, highlighting different emitting CR
populations, is regarded as the main piece of evidence that CRs below $10^{15}$~eV are Galactic in
origin. The only non-AGN galaxy detected before the LAT era was the nearby Large Magellanic Cloud.
The LAT has so far detected \g-ray emission from galaxies of the local group, the two
Magellanic clouds
\cite{ref:lmc,ref:smc} and the Andromeda galaxy \cite{ref:m31}, and from some nearby starburst
galaxies \cite{ref:starburst}.

Thanks to the improved angular resolution of the LAT, the Large Magellanic Cloud was the first
external galaxy ever resolved in high-energy \g-rays \cite{ref:lmc}, leading to the first map of CR
acceleration and propagation in a galaxy. The \g-ray emission correlates with massive-star formation
rather than interstellar gas. This reinforces the idea that the energy for accelerating CRs in
galaxies
is provided by massive stars, thanks to the catastrophic explosions taking place at the end of their
lives and, perhaps, to the collective action of stellar winds in massive-star clusters. On the
other hand, the lack of correlation with the distribution of gas would require propagation lengths
shorter than usually assumed in the Milky Way.

The link between the acceleration of particles and massive stars is further supported by the
correlation found between the \g-ray luminosity and the massive-star formation rate, see
Fig.~\ref{fig:galaxies}.
This correlation is not fully understood yet, but might be analogous to the correlation found
between radio and infrared luminosities of galaxies. An enlarged sample of external galaxies studied
by the LAT will allow us to use this correlation to evaluate the contribution by
unresolved non-active galaxies to the isotropic diffuse \g-ray emission.

\subsection{The isotropic diffuse \g-ray emission}

The LAT measured the spectrum of the isotropic diffuse \g-ray emission and found it to be
consistent with a
featureless power low of index $\sim 2.4$ in the energy range 200 MeV-100 GeV \cite{ref:egb}. This
is expected to be largely due to populations of unresolved extragalactic sources. The population
synthesis of AGN detected by the LAT (see \ref{agn}) constrains their contribution to be $<30\%$
\cite{ref:AGNpop}. The contribution from non-active galaxies is under evaluation, for the first time
based not only on theoretical models but also on a sample detected by the LAT (see \ref{extgal}).

Yet, at present there is still room for
\emph{truly diffuse} extragalactic emission, which can be due to many different processes, like
large-scale structure formation, interactions of ultra-high-energy CRs with
the extragalactic low-energy background radiation, annihilation or decay of cosmological
DM, e.g. \cite{ref:egbdm}. On the other hand, interactions of CRs with debris at the
outer frontier of the solar
system might partially contribute to the spectrum of the isotropic diffuse
emission \cite{ref:moskalenko2009}.

\section{Indirect Dark Matter searches in \g-rays}\label{sec:dm}

One of the major open issues in our understanding of the Universe is the existence of an
extremely-weakly interacting form of matter, DM, supported by a wide range
of observations including large scale structures, the cosmic microwave background and the
isotopic abundances resulting from the primordial nucleosynthesis. 
Complementary to \emph{direct} searches being carried out in underground
facilities and at accelerators, the \emph{indirect} search for DM
is one of the main items in the broad \F{} Science menu.

The word ``indirect'' denotes here the search for signatures of Weakly
Interactive Massive Particle (WIMP) annihilation or decay processes through the final
products (\g-rays, electrons and positrons, antiprotons) of such processes.
Among many other ground-based and spaceborne instruments, the LAT plays a prominent role in this
search through a variety of distinct
search targets: \g-ray lines, Galactic and isotropic diffuse
\g-ray emission, dwarf satellites, CR electrons and positrons (for the latter item
see Section~\ref{sec:cr}).

\subsection{\g-ray lines}
The quest for a possible narrow line in the diffuse \g-ray emission
arises naturally since photons can be produced in two-body DM particle
annihilations $\chi\chi \rightarrow \gamma X$ or decays
$\chi \rightarrow \gamma X$.
Since, in most scenarios, DM particles are electrically neutral (and therefore
do not couple directly to photons) such processes only occur at high
orders and the branching ratios are expected to be strongly suppressed.
Despite the subsequent weakness, a photon line, if present, is easy to identify
and distinguish from the standard astrophysical sources of \g{} rays---whose
flux is dominant in most situations. Therefore, this discovery channel
features a distinctive
experimental signature that, if observed, would incontrovertibly indicate
new physics at work.

\begin{figure}[!hbtp]
  \begin{center}
    \includegraphics[width=0.75\textwidth]{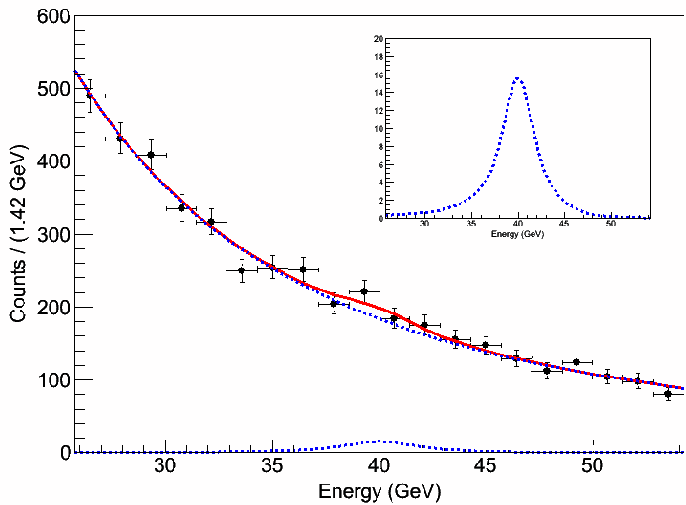}
    \caption{Binned representation of the fit procedure (here centered at
      40~GeV) used to derive the upper limit on the flux of a possible photon
      line contribution in the all-sky (except for part of the Galactic plane)
      spectrum (from \cite{ref:dmlines}).
      The two dotted lines represent the background (modeled with a
      power law) and the signal from the fit, respectively, while the red line
      is their sum. The insert shows a close-up of the instrument response
      to a monochromatic line at 40~GeV, which is used to model the signal.}
    \label{fig:dmline}
  \end{center}
\end{figure}

The detector response to a monochromatic line is not a
monochromatic line and the effect of the finite energy resolution cannot
be ignored. The response can be modeled by means of Monte Carlo
simulations (see the insert in Figure \ref{fig:dmline}) and verified with tests
at accelerators so that it can be effectively folded into the procedure and used
to asses the statistical significance of a possible line component in the
measured count spectra.

No significant evidence of \g-ray line(s) has been found in the
first 11 months of data, between 30 and 200 GeV~\cite{ref:dmlines} and
work is ongoing to extend the energy range of the analysis and include more
data. The detailed discussion of the upper limits obtained and their relevance
in the context of specific DM models is beyond the scope of this brief overview.

\subsection{DM signals from the Galactic halo}
The Milky Way halo and the Galactic center are obvious candidates for
indirect dark matter searches in \g-rays: given the large DM content, a
large annihilation signal can be potentially expected. The main challenge is presented by the strong
\g-ray foreground comprising the Galactic diffuse emission (cfr. section~\ref{sec:gde}). Indeed, the
detailed modeling of this foreground is currently the main limiting factor
for DM searches in this channel.

In the case of the Galactic center a firm assessment of the conventional
astrophysical signals is furthermore complicated by the problem of source
confusion, due to the limited angular resolution of the instrument,
and of pile-up along the line of sight.

\subsection{Dwarf galaxies}

Dwarf satellites of the Milky Way are
among the cleanest targets for indirect dark matter searches in \g-rays.
They are systems with a very large mass/luminosity ratio
(i.e. systems which are largely DM dominated).
The LAT detected no significant emission from any of such systems and the upper
limits on the \g-ray flux allowed us to put very stringent constraints
on the parameter space of well motivated WIMP models~\cite{ref:dmdwarfs}.

A combined likelihood analysis of the 10~most promising dwarf
galaxies, based on 24~months of data and pushing the limits below the thermal
WIMP cross section for low DM masses (below a few tens of GeV), is
currently under preparation and will be published soon.

\section{Direct cosmic-ray measurements}\label{sec:cr}

The electron component of the primary CR radiation is widely
recognized as a unique probe to address a number of significant questions
concerning the origin of CRs and their propagation in our galaxy
(see \cite{ref:crereview} for a synthetic review).
Several different experiments have recently published new data of
unprecedented quality and spanning energy ranges never explored before,
stirring up the interest of the scientific community, mostly in connection
with the possible indication of the existence of an extra component in the
energy spectrum, which is not part of the standard CR
paradigm.

\subsection{The all-electron spectrum}

The LAT intrinsic capability for detecting high-energy CR electrons (and positrons)
was already recognized by the collaboration in the early stages of the
instrument development. This was demonstrated with the publication of
the first high-statistics spectrum of CR $e^+ + e^-$ between 20 GeV and
1 TeV, based on the data from the first six months of the
mission~\cite{ref:crespectrum}.
This capability derives from the unique combination of large acceptance and
long observation time; this combination results in an effective exposure
factor of the order of a few $10^8$~m$^2$~sr~s at 100 GeV (including all the
effects of the event selections and instrument dead time and duty cycle),
which is significantly larger than any other experiment ever flown to measure
the electron component of the cosmic radiation in the LAT energy range.

\begin{figure}[!hbtp]
  \begin{center}
    \includegraphics[width=0.75\textwidth]{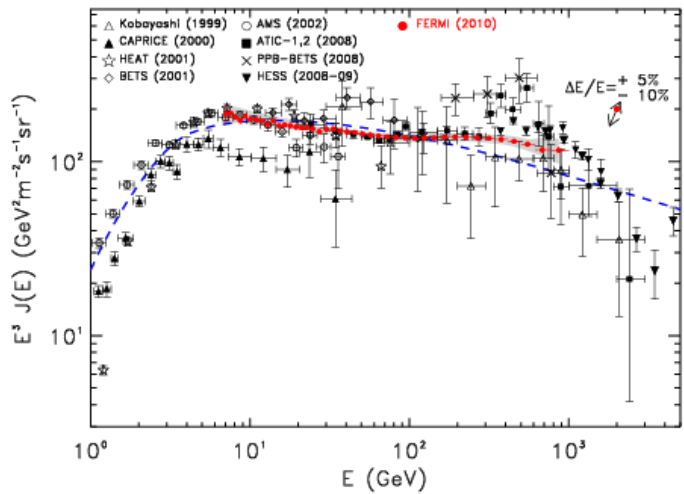}
    \caption{CR electron spectrum measured by the \F~LAT
      (red points, from~\cite{ref:crespectrum2}). The gray shaded band
      indicates the systematic uncertainties associated with the flux values,
      while the blue dashed line represents the prediction of a diffusive
      propagation model tuned on the pre-\F\ data.
      Many other recent measurements are included for completeness.}
    \label{fig:crespectrum}
  \end{center}
\end{figure}

Figure~\ref{fig:crespectrum}
shows the updated $e^+ + e^-$ spectrum, based
on the first year of data, published in~\cite{ref:crespectrum2}. It extends the
previous result~\cite{ref:crespectrum} down to 7~GeV---the lowest accessible
energy for primary electrons and positrons in the \F\
orbit (with an inclination of~$25.6^\circ$).
The spectrum does not show any evidence for a sharp spectral feature, such as
the one reported by ATIC~\cite{ref:creatic}. The new low-energy data points, though,
exacerbate the tension with the hypothesis of a single power law spectrum,
which could not be ruled out in~\cite{ref:crespectrum}.

In fact, several different authors showed that the LAT $e^+ + e^-$ spectrum,
along with the positron fraction measured by the PAMELA
experiment~\cite{ref:pospamela}, can be well fitted with the addition of a
separate high-energy $e^+/e^-$ component, whose origin might be due to
nearby pulsars or annihilation of WIMP dark matter.
Other authors, like \cite{ref:creblasi} and \cite{ref:crewaxman}, however, point to
alternative explanations of the \emph{lepton excesses} not invoking any extra-components.

\subsection{Anisotropies in CR electrons}

Thanks to its large exposure factor, and to it being an all-sky instrument,
\F\ offers a unique opportunity for the measurement of possible anisotropies
in the arrival directions of CR electrons and positrons, that could
potentially provide important information (complementary to the energy
spectrum) about their origin.

With more than 1.6 million candidate electrons above 60 GeV in the first year
of operation, we found no evidence for anisotropies on any angular scale.
The search was performed by means of integrated skymaps with different regions
of interest (with radii ranging from $10^\circ$ and $90^\circ$) and though
a spherical harmonic analysis.
For a dipole anisotropy---which is an interesting case study in the situation
in which a single nearby source dominates the high-energy electron
spectrum---the LAT upper limits range from 0.5\% to 10\%, depending on the
energy range. This is close to the level where we might indeed expect some
signal and it will get better as new data are analyzed.

\section{Final remarks: \F{} and Astroparticle Physics}

The \F{} observatory provides a paradigmatic example of the fruitful interplay between Astrophysics
and Particle Physics, which has been taking place over the last decades.
One the one hand, the highlight results we presented show how the significant observational advances
of
the \F{} era are largely due to the technological developments coming from accelerator-based
detectors, notably in this case to silicon tracking devices.

On the other hand, \F{} is contributing to a deeper understanding of high-energy processes occurring
in the Universe, often much more energetic than those produced by our Earth-based particle
accelerators. High-energy
particles of cosmic origin played an important role in the birth of subnuclear Physics and are now
regarded as a fundamental component of galaxies, as well as messengers of possible exotic
phenomena beyond the standard model of Particle Physics. The LAT is mapping with unprecedented
accuracy the high-energy facet of the sky,
performing direct measurements of the charged cosmic radiation and searching for signatures
of Physics beyond the standard model. Further advancements are foreseen in next years thanks to the
complementarity with other multi-wavelength/messenger Astroparticle detectors and accelerator-based
experiments.

\acknowledgments
The \F{} LAT Collaboration acknowledges support from a number of agencies and
institutes for both development and the operation of the LAT as well as
scientific data analysis. These include NASA and DOE in the United States,
CEA/Irfu and IN2P3/CNRS in France, ASI and INFN in Italy, MEXT, KEK, and
JAXA in Japan, and the K.~A.~Wallenberg Foundation, the Swedish Research
Council and the National Space Board in Sweden. Additional support from INAF
in Italy and CNES in France for science analysis during the operations phase
is also gratefully acknowledged.

\end{document}